# Porosity Effects on Cyclic Gas Invasion and Trapping in Deformable Porous Media

Haiyi Zhong [a], Jieting Long [b], Xiangyu Ding [a], Zhongzheng Wang [c†], Yixiang Gan [a†]

[a] School of Civil Engineering, The University of Sydney, NSW 2006, Australia

[b] School of Computer Science, The University of Sydney, NSW 2006, Australia

[c] School of Mechanical, Medical and Process Engineering, Faculty of Engineering, Queensland University of Technology, QLD 4001, Australia

†Corresponding authors: Zhongzheng Wang: zhongzheng.wang@qut.edu.au; Yixiang Gan: yixiang.gan@sydney.edu.au

Key points:

1. Initial porosity controls the gas invasion mode, shifting from cavity expansion to mixed cavity-fingering and localised pore invasion.
2. Pressure evolution during cyclic injection reflects both pore-scale events and cycle-to-cycle gas distribution.
3. Cyclic injection increases residual trapping and interfacial complexity, with lower porosity yielding higher trapping and rougher clusters.

## Abstract

Fluid transport in deformable porous media is central to many biophysical and geophysical processes. While extensive studies exist, how porosity governs fluid behaviour in deformable systems during cyclic injection remains elusive. Here, we investigate gas-liquid multiphase flow in a quasi-2D Hele-Shaw cell packed with soft hydrogel particles at different initial porosities. Alternative gas and water injection experiments, combined with high-resolution imaging and continuous pressure monitoring, are used to quantify gas dynamics and pressure evolution. Results show that the gas entry pressure increases as porosity decreases, consistent with a Young-Laplace estimation based on effective pore-throat width. After entry, invasion shifts from cavity-dominated expansion in high porosity packings to localised pore invasion in low porosity packings, with a mixed cavity-fingering regime at intermediate porosity. Pressure fluctuations are linked to pore-scale gas escape and internal gas redistribution. Low porosity packings produce frequent small-amplitude pressure drops, whereas higher porosity packings produce more discrete pressure relaxations. Across cycles, the decreasing mean pressure suggests preferential-pathway reuse and reduced local capillary constraints. Residual gas saturation increases systematically with injection cycles and reaches higher terminal values as porosity decreases. Specific interfacial length increases as available pore space decreases and follows a power-law relationship with gas cluster size, with scaling exponent decreases as porosity decreases and cycling progresses. Together, these results demonstrate that gas trapping in deformable porous media depends on both initial packing structure and cyclically evolving gas-solid interactions. This study provides insights for interpreting porosity-dependent trapping and reinvasion during repeated gas injection.

## Plain language summary

Porous materials in the subsurface can store gases such as carbon dioxide and hydrogen, but many of these materials can change shape when fluids move through them. These changes affect where gas travels, when it escapes, and how much remains trapped after injection. We studied this process using a transparent cell filled with water and soft hydrogel grains. Gas and water were injected repeatedly through packings with three different porosities, meaning different amounts of open space between grains, while images and pressure data were recorded. The results show that denser packings, with less open space, require higher pressure before gas can enter. They also make gas move through narrower, more localised pathways, causing smaller but more frequent pressure drops. Looser packings allow gas to grow as larger cavities and release in larger events. Over repeated injection cycles, more gas becomes trapped in all packings, but denser packings retain the most gas and form more fragmented gas shapes. These findings show that the initial pore structure and its evolution during repeated injection strongly control gas storage and release in soft porous materials. The results help interpret gas trapping and reinvasion during underground gas storage.

# 1 Introduction

Fluid transport in porous media governs a broad range of biophysical and geophysical processes. In subsurface applications, multiphase displacement controls the storage efficiency and security during geological carbon dioxide sequestration and underground hydrogen storage (Krevor *et al*., 2023; Zhong *et al*., 2024). In biological contexts, it regulates oxygen and nutrient delivery within porous scaffolds, directly influencing cell survival and function (Op Den Buijs *et al*., 2009; Ferroni *et al*., 2016). Across these diverse applications, understanding how fluids redistribute, become trapped, and evolve over time is essential for both performance and risk management.

Many pore-scale insights have been obtained in rigid porous media, where pore geometry is fixed during flow (Krevor *et al*., 2012; Zuo and Benson, 2014; Wang *et al*., 2022). However, subsurface formations and many biological materials can deform in response to fluid pressure changes (Op Den Buijs *et al*., 2009; Qiu *et al*., 2023). Such deformation modifies pore size, throat apertures, and preferential pathways, producing displacement behaviours that differ from those in rigid media. Controlled and optically accessible experiments remain challenging because observable deformation generally requires either high driving pressure or very soft, index-matched materials (MacMinn *et al*., 2015). Transparent hydrogel grains provide a useful model system for this purpose; for example, for example, Zadeh *et al.* (2023) showed that grain elasticity can redirect flow pathways and alter deformation during fluid injection. Despite these recent advances, links among porosity, gas trapping, and pressure signatures in deformable porous media remain insufficiently resolved.

Porosity is a state variable that reflects packing preparation and mechanical confinement, and it is closely associated with both hydraulic and poromechanical responses. A looser, higher porosity packing generally permits larger particle rearrangement and matrix deformation under injection pressure, favouring cavity expansion. A denser, lower porosity packing provides stronger mechanical confinement and is more likely to promote localised pore invasion and snap-off behaviours consistent with capillary-dominated migration in compacted granular media (Sandnes *et al.*, 2011; Lee *et al.*, 2020). Collectively, these findings underscore the central role of porosity in affecting mass transfer. However, the influence of porosity on quantitative trapping characteristics and the associated pressure signatures have not yet been systematically investigated for deformable porous media.

Cyclic fluid flow introduces additional complexity to these poromechanical interactions and can markedly alter transport behaviour. In geophysical applications, carbon dioxide and hydrogen storage operations commonly involve repeated injection and withdrawal cycles (Liebscher *et al.*, 2013; Krevor *et al.*, 2023). However, most laboratory studies have focused primarily on a single drainage-imbibition cycle (Krevor *et al.*, 2012; El-Maghraby and Blunt, 2013; Zuo and Benson, 2014). The limited studies involving sequential cycles remain inconclusive: some report that residual gas saturation evolves progressively with successive injection cycles (Herring *et al.*, 2016; Abdoulghafour *et al.*, 2020; Herring *et al.*, 2021; Wang *et al.*, 2023), whereas others observe negligible changes over time (Ruprecht *et al.*, 2014; Saeedi and Rezaee, 2012; Garing and Benson, 2019). Moreover, cyclic fluid flow is also accompanied by oscillatory pressure variations, which are known to strongly influence mass transport (*e.g.*, oxygen and solute) in biophysical systems by promoting fluid motion (Op Den Buijs *et al.*, 2009) and restructuring pore connectivity (Ferroni *et al.*, 2016). Together, these findings indicate that cyclic injection can alter the transport properties of porous media, yet its consequences in deformable granular systems remain largely unknown.

To address aforementioned research gaps, this study carried out cyclic gas-liquid displacement experiments in vertical Hele-Shaw cells containing quasi-2D monolayer hydrogel particles, which allows for clear visualisation and quantitative analysis of the trapped gas. Three initial porosity states are examined to determine how packing porosity is associated with (i) capillary entry pressure and initial invasion mode, (ii) pressure signatures during drainage, and (iii) residual saturation, cluster morphology, and specific interfacial length over repeated cycles. Finally, we discuss the limitations of the current work and outline directions for future investigation.

# 2 Methodology

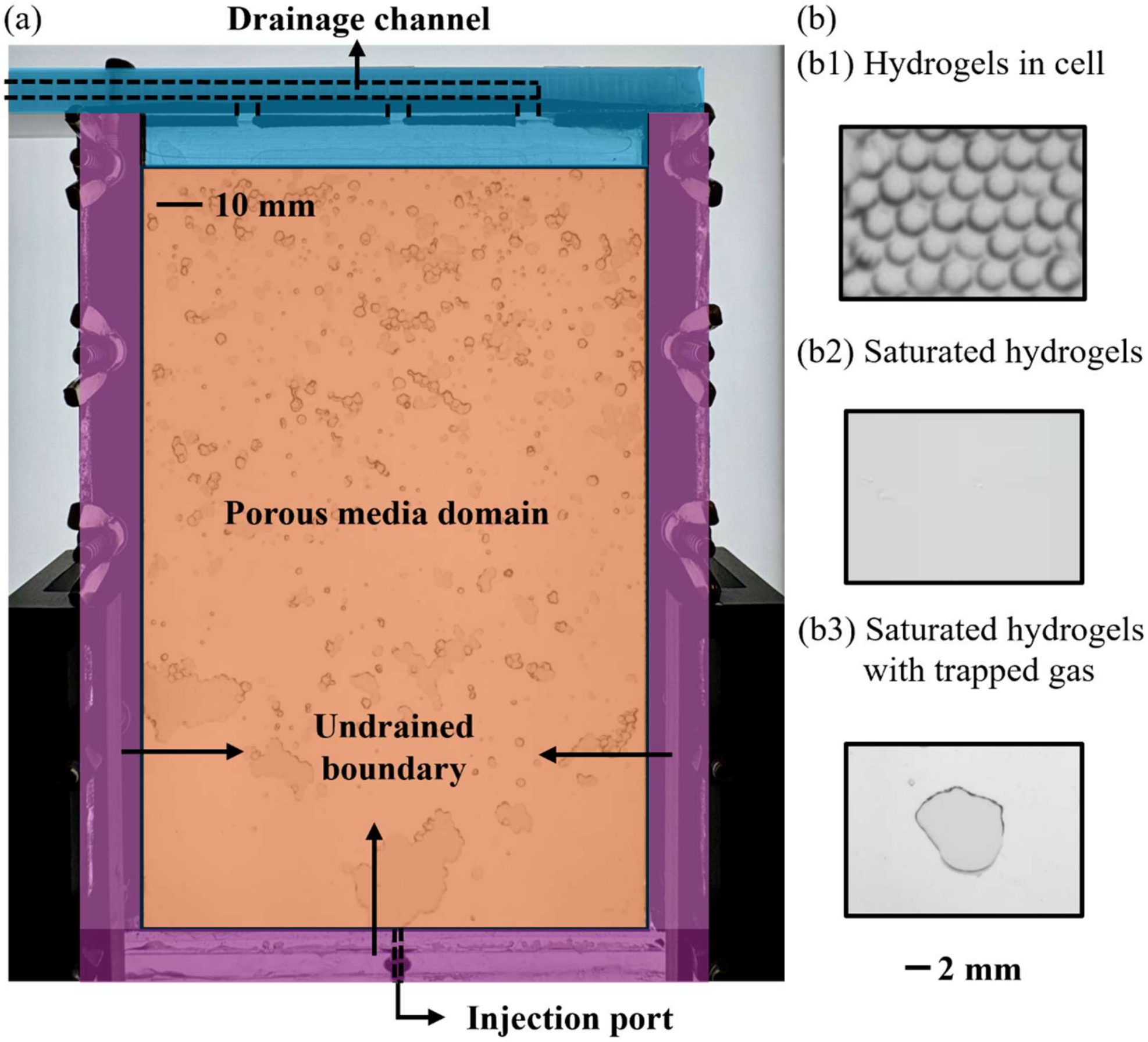

**Figure 1 Experimental setup.** (a) Hele-Shaw cell and boundary conditions. Side and bottom boundaries were sealed, while the top module contained a drainage/vent channel; (b) Different hydrogel conditions within the Hele-Shaw cell.

### 2.1 Experimental system

In this study, polyacrylamide hydrogel beads (JRM Chemical) were used to assemble the granular media. The beads were supplied in a dry granular form and swelled into soft grains upon hydration, as shown in Figure 1 (b1). The terminal polymer fraction in each bead is less than 1%, giving a refractive index nearly identical to that of water. As a result, the water-saturated hydrogel packing is almost transparent (Figure 1 (b2)), enabling trapped gas to be visualised clearly (Figure 1 (b3)). The swollen hydrogel grains have a measured density ~1.02 g/cm$^3$ and a Young's modulus of 29.8 ± 3.3 kPa measured following the experimental setup described by Zhao *et al*., 2026. The as-received product contained grains of variable sizes, and a staged sieve procedure was conducted to improve size uniformity. First, dry grains were sieved to 0.25-0.3 mm. After 24 hours of swelling in demineralised water, the grains expanded to 1.7-2.1 mm. To further enhance uniformity, they were then sieved with customised 3D printed sieves (1.9- and 2.1-mm openings), and only grains within this range were retained.

Experiments were conducted in a Hele-Shaw cell constructed from two transparent acrylic plates (120 × 160 × 7.5 mm) separated by acrylic side strips (10 × 160 × 1.5 mm) lined with 1 mm EPDM foam. This configuration produced a gap of 2.15 ± 0.02 mm, forming a quasi-2D particle monolayer while preventing leakage. The top of the cell was fitted with a 3D printed module serving as a drainage or vent channel. The bottom plate incorporated an injection port connected to syringe pumps via plastic tubing. After assembly, all edges were sealed with silicone adhesive. The overall experimental configuration is shown in Figure 1 (a).

Before each test, the cell was saturated with deaired water and pre-weighted particles were packed to achieve the target porosity. Once the packing phase was completed, the cell was mounted on an optical platform (flatness ±0.05 mm over 0.36 m²) and illuminated with an LED backlight (JSIONX, JS-DBL209-318). Fluids were injected from the base at a constant rate $Q$ of 3 mL/min using syringe pumps (Y.H. POWER, QHLR-0140). During imbibition and drainage processes, the initial capillary number $Ca = \frac{\mu \cdot U}{\gamma}$ is 3.23×10$^{-6}$ and 5.81×10$^{-8}$, respectively, indicating that both displacement processes occur in the capillary-dominated regime (Lenormand et al., 1988). Here, $\mu$ is the viscosity of the injected fluids, $U$ is the Darcy velocity estimated from the imposed flux as $\frac{Q}{Wb}$ ($W$ is the cell width and $b$ is the cell gap) and $\gamma$ is the surface tension. To assess gas compressibility, we used the compressibility number C= $\frac{12\mu Q V_g(0)}{b^4 W^2 p_{atm}}$ is $2.2 \times 10^{-5}$, suggesting that compression-induced unsteady displacement is not expected to dominate the overall flow dynamics (Cuttle and MacMinn, 2023). Here, $V_g(0)$ is the initial gas volume (*i.e.*, 200 ml in this study). The bond number $B_O = \frac{\Delta\rho \cdot g \cdot L^2}{\gamma}$ for all cases is around 0.6. Here, $\Delta\rho$ the density difference between the gas and liquid phases, $g$ is the gravitational acceleration and $L$ is the characteristic length, which in this case refers to the gap distance of the Hele-Shaw cell. Gas migration was imaged with an industrial digital camera (DAHENG, MER2-2000-19U3M) at 4 frames per second and 5,496 × 3,672 pixel with a spatial resolution of 2.4 $\mu$m/pixel. Injection pressure was monitored using a pressure transducer (MEACON, SUP-P3-000D) with data logged by a MEACON MIK-RN 3000 system, with a pressure accuracy of 0.001kPa.

For each test, four imbibition-drainage cycles were performed, each consisting of a 5-minute air injection and a subsequent 5-minute water injection. The injection duration was selected given that a steady state displacement process can be reached. A total of three porosity conditions were tested, including high porosity (φ = 0.346 ± 0.027), medium porosity (φ = 0.299 ± 0.019), and low porosity (φ = 0.252 ± 0.020). Each condition was tested in triplicate to validate repeatability, resulting in a total of nine tests. Freshly prepared hydrogel grains were used for each test to avoid irreversible deformation or aging effects.

**2.2 Image processing and bubble dynamics analysis**

Bubble detection and tracking were implemented in Python using OpenCV-based preprocessing and the Segment Anything Model 2 (SAM2; Ravi *et al.*, 2024). Prior to SAM2 segmentation, key frames were pre-processed using computer-vision-based methods to enhance image quality and improve bubble identification. For each image, a plain background frame was first subtracted to correct for illumination non-uniformity, followed by Gaussian smoothing and normalisation using a dynamic light-field matrix to minimise residual brightness gradients. The resulting grayscale images were segmented using a histogram-based adaptive threshold. Morphological filtering was then applied to remove noise and small artifacts. The remaining connected regions were treated as candidate bubble masks, which were used as positive prompts for SAM2. The segmentation was performed using a chunked propagation strategy, in which mask propagation was carried out over shorter temporal segments rather than over the full sequence at once. This strategy was adopted to tackle the issues of error accumulation in the memory-attention module within SAM2, especially during the long-sequence propagation due to the complex bubble shapes and low image contrast occurred in our experiments. For frame-wise analysis, each propagated mask was decomposed into connected components so that spatially disconnected bubbles were represented as individual objects. Inter-frame association was established based on spatial overlap. The connected components smaller than a prescribed area threshold (corresponding to the projected area of a single particle) were removed to focus on the dynamic of major bubbles.

Each bubble was assigned a persistent identifier to enable temporal tracking and the identification of dynamic events, including bubble translation, splitting, merging, and escape, as demonstrated in Figure 2. During snap-off or coalescence events (Figure 2 (a1)-(a2)), the original bubble identifiers are removed and newly generated bubbles are assigned new identifiers. During mobilisation, the bubble identifier remains unchanged between consecutive frames, as shown in Figure 2 (b1)-(b2). For escape events, the corresponding bubble identifier was removed after the bubble leaves the porous domain. It should be noted that an escape event may occur simultaneously with splitting or merging events. As shown in Figure 2 (b1)-(b2), bubble escape is accompanied by snap-off. In such cases, the effective escaped volume is determined from the net reduction in total gas area before and after the event (*i.e.*, area[#b4-#b8-#b9-#b10]).

For each tracked bubble, the area and geometric descriptors were quantified at each frame. Bubble velocity was estimated from centroid displacement between consecutive frames. This velocity was calculated only when the same bubble identifier persisted across two consecutive frames. If a splitting, merging, or escape event occurred between two frames, the original identifier was no longer present in the subsequent frame, and no velocity was assigned for that transition.

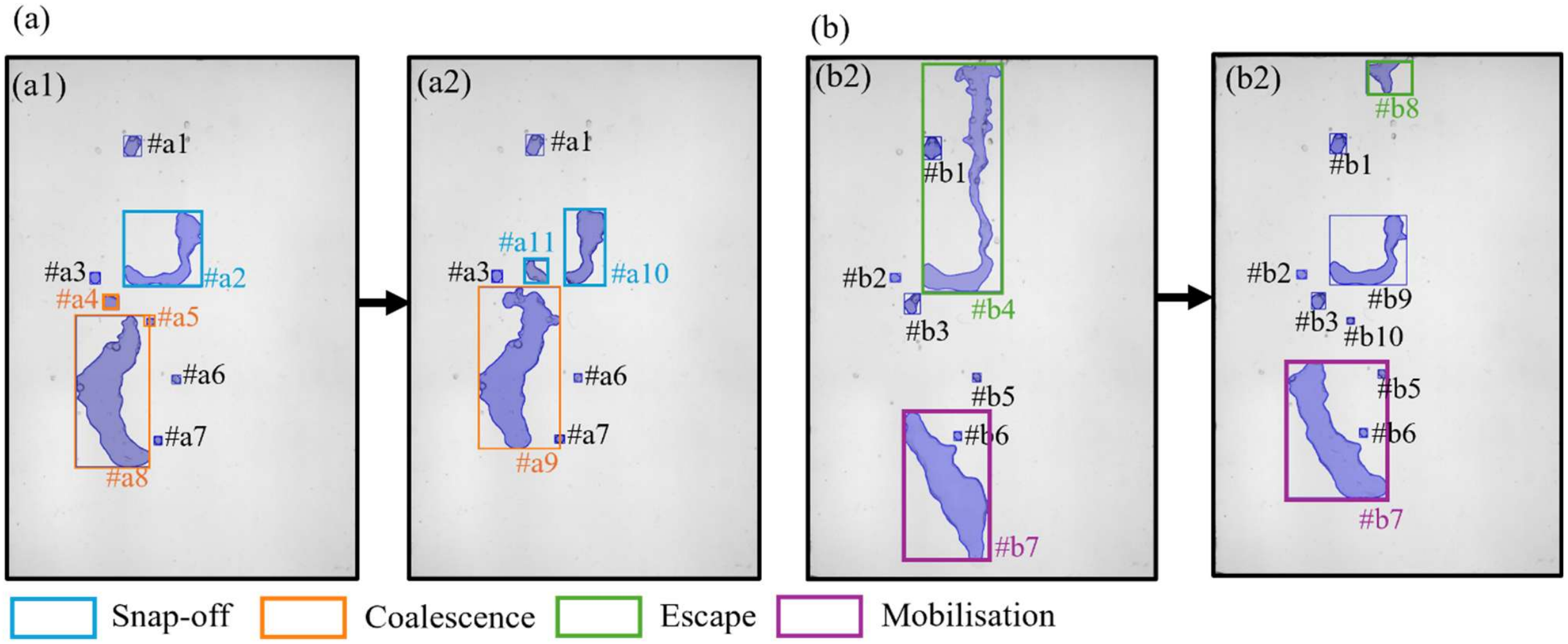

**Figure 2 Identification of bubble dynamic events.** (a) Snap-off and coalescence events; (b) Escape and mobilisation events. Each bubble is labelled with its corresponding identifier, with ongoing events highlighted. The blue, orange, green and purple boxes denote snap-off, coalescence, escape and mobilisation events, respectively.

# 3 Results and Discussion

### 3.1 Initial gas dynamics and pressure signatures

*3.1.1 Prior to gas percolation*

The initial gas injection dynamics before percolation were interpreted from the hydrostatic-corrected injection pressure, $P_{inj}^{*} = P_{in} - P_{h}$, where $P_{in}$ is the gas pressure measured at the inlet and $P_{h} = \rho_{w} g H$ is the hydrostatic pressure exerted by the overlying water column of height $H$. At the onset of invasion, the characteristic gas velocity at the inlet is approximated by the imposed injection velocity, consistent with the quasi-static and weakly compressible conditions of the system. Figure 3 (a) shows the smoothed apparent injection pressure evolution using 5-point centred moving average, with $T_{E} = 0$ marks the onset of gas invasion. The pressure at this moment, defined as the apparent entry injection pressure $P_{inj,entry}^{*}$ can be expressed as $P_{inj,entry}^{*} = P_{c,entry} + P_{v} + P_{i}$, where $P_{c,entry}$ is the capillary entry resistance, $P_{v}$ is the viscous pressure loss, and $P_{i}$ is the inertial contribution. Order-of-magnitude estimates indicate that viscous and inertial effects are relatively small under the present experimental conditions. The tube viscous pressure loss estimated from Poiseuille flow is $P_{v,t} \approx 9.4 \times 10^{-5}$ kPa, while the Reynolds number $Re_t \approx 1.7$ confirms laminar flow and negligible inertial effects. In the porous medium, Darcy's law estimates give $P_{v,\mathrm{pm}} \sim 0.012$-$0.044$ kPa, while the pore-scale Reynolds number $Re_{\mathrm{pm}} \approx 0.38 < 1$ indicates negligible inertial contributions. These contributions are at least an order of magnitude smaller than the observed entry pressure (approximately 0.6-1.0 kPa). Therefore, to leading order, $\mathrm{P}^{*}_{inj,entry} \approx P_{c,\mathrm{entry}}$.

The capillary entry threshold is controlled by the larger of the pressure required for gas to enter the injection port $P_{\mathrm{orifice}}$ and that required to enter the hydrogel pore space $P_{\mathrm{pore}}$. For gas invading a water-saturated pore space, the capillary pressure follows the Young-Laplace relation $P_c = \gamma\left(\frac{1}{R_1} + \frac{1}{R_2}\right)$, where $R_1$ is associated with the in-plane pore-throat curvature, and $R_2$ represents the out-of-plane curvature constrained by the Hele-Shaw gap. Thus, the pore-entry pressure was approximated as:

$$P_{pore} = \gamma\left(\frac{2\cos\theta}{w_{\mathrm{eff}}} + \frac{2\cos\theta}{b}\right), \qquad (1)$$

where $\theta$ is the contact angle, $b = 2.15\,mm$ is the cell gap, the air-water interface tension $\gamma = 0.072 N/m$ and $w_{\mathrm{eff}}$ is the effective pore-throat width, which can be estimated from the porosity-based projected particle geometry, with details provided in Appendix A. The hydrogel-water-air system was treated as strongly water-wet, and θ = 0° was used in the Young-Laplace estimate.

The injection-port threshold was estimated as:

$$P_{\mathrm{c,orifice}} = \frac{4\gamma\cos\,\theta}{d_{\mathrm{port}}}, \qquad (2)$$

where $d_{\mathrm{port}} = 1$ mm. $P_{\mathrm{port}} \approx 0.288\,kPa$, which is smaller than $P_{\mathrm{pore}}$ for all cases. Therefore, the measured entry pressure is primarily controlled by gas entry into the hydrogel pore network rather than by the injection port.

The inset of Figure 3 (a) compares the measured entry pressures with the Young-Laplace estimates, showing reasonable agreement and the same increase in entry pressure as porosity decreases. Once the entry threshold was exceeded, distinct invasion patterns emerged at $T_E$ (Figure 3 (b1)). In the low porosity case, the rapid localised pore invasion resembles a Haines-jump-like capillary depinning event, producing the fastest gas-front advance. In contrast, the high and medium porosity cases show cavity expansion near the inlet. This porosity-dependent post-entry behaviour leads to distinct initial migration modes. Figure 3 (b2) presents the gas invasion morphologies at a similar gas volume (invaded pore volume of approximately 18%). The high porosity system exhibits cavity-dominated fluidisation. The low porosity system shows immediate pore-scale invasion through localised pathways. The medium porosity case displays a mixed mode, with cavity formation near the inlet followed by localised front fingering. This porosity-dependent transition is consistent with previous experimental observations by Lee *et al.* (2020).

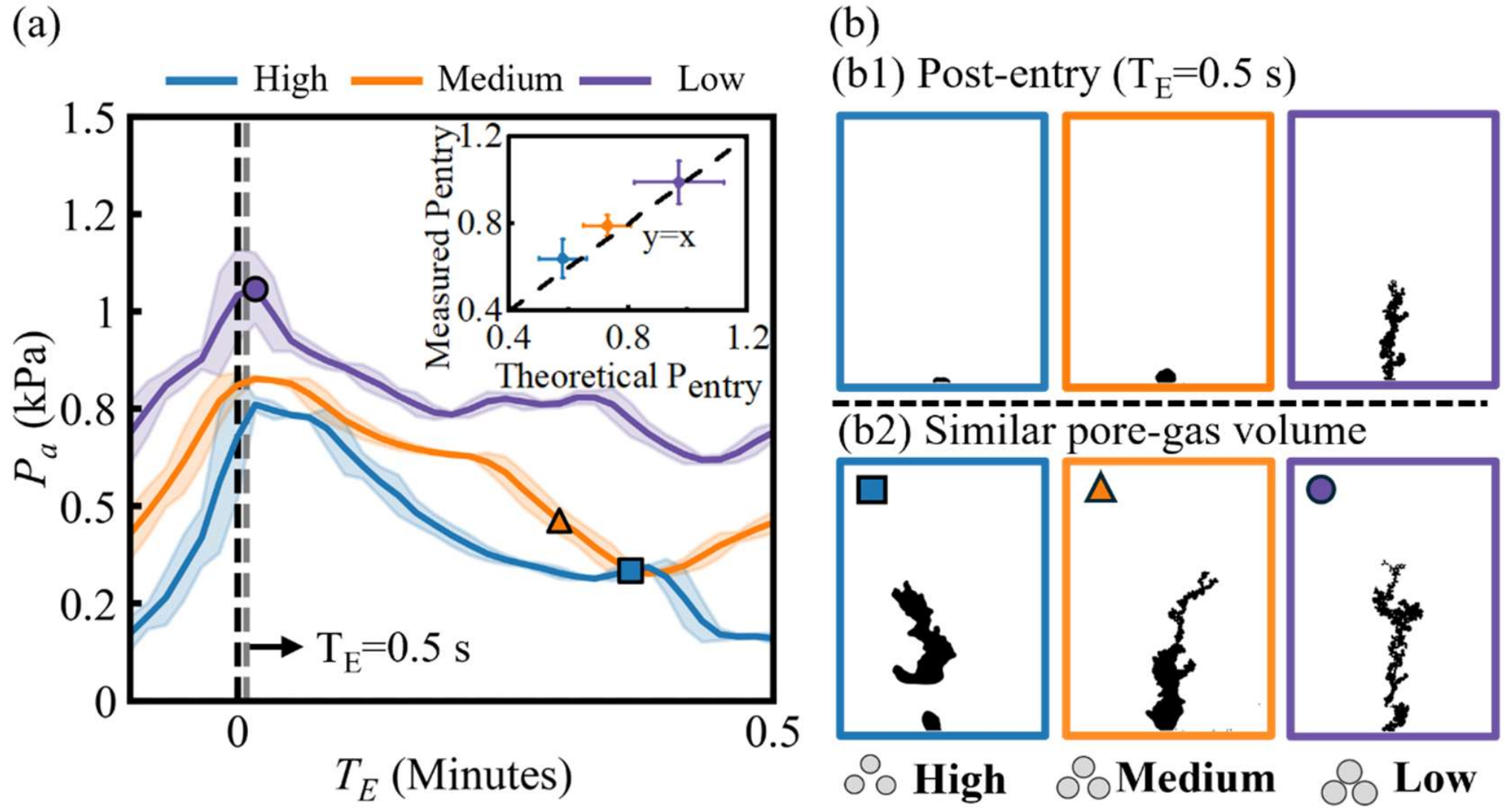

**Figure 3 Gas injection dynamics prior to gas percolation.** (a) Evolution of apparent injection pressure (The shaded regions indicate the error bands and the lines represent the mean values smoothed using a moving average with a window size of 3 seconds. The inset compares the measured and theoretical critical entry pressures. Error bars denote experimental uncertainty arising from variations in cell gap, particle size, and porosity for the theoretical estimates, and from the pressure variation within $T_E$ ±2 s for the measured values.); (b) Typical gas injection patterns: (b1) At $T_E$=0.5 s; (b2) At similar segmented in-cell gas fraction (Black indicates gas and white indicates water and hydrogel particles. The corresponding time and pressure for the corresponding patterns are denoted as square (high porosity), triangular (medium porosity), circular (low porosity) markers in (a)).

*3.1.2 First drainage period (Air 1)*

After gas entry, the subsequent evolution of $P^{*}_{inj}$ exhibits porosity-dependent fluctuations associated with intermittent gas release from the porous domain (Figure 4 (a)). To investigate this coupling, the pressure signal during the first drainage period (Air 1) was smoothed using a 3-second moving average and compared with individual gas escape events. Escaped gas volume was estimated from the escaped gas area multiplied by the cell gap. A clear temporal correlation is observed between pressure drops and gas escape events, indicating the rapid gas release is the primary driver of pressure relaxation. Specifically, the pressure fluctuation pattern depends on both the frequency and magnitude of escape events. The low porosity case shows frequent escape events with relatively small volumes, producing repeated small-amplitude pressure drops and a highly intermittent pressure response. In contrast, the high and medium porosity cases show less frequent escape events with larger gas volumes, producing more discrete pressure relaxations. The probability density distribution of escaped gas volume, also shown in Figure 4 (a), supports this interpretation: the low porosity case distribution is concentrated at smaller volumes, whereas the higher porosity cases extend toward larger release volumes. These observations suggest that lower porosity favours fragmented release through localised pathways, while higher porosity permits accumulation and escape of larger gas clusters or cavities.

Pressure fluctuations are not controlled solely by gas escape events at the outlet, and they may also be influenced by internal pore-scale gas redistribution. To access the potential contribution of transient bubble motion, Figure 4 (b) compares the normalised maximum bubble velocity during mobilisation events with the normalised bubble size. Here, $V_{max}$ is the maximum centroid-based bubble velocity measured between two consecutive frames, $V_{inj}$ is the imposed gas injection velocity, $A_{max}$ is the maximum projected bubble area over the same two-frame interval, and $A_{ref}$ is the projected area of the porous domain within the Hele-Shaw cell. Although the nominal capillary number is low, individual bubbles can reach velocities more than two orders of magnitude higher than the nominal injection velocity during rapid local rearrangements. This observation suggests that local pore-scale displacement may temporarily depart from the nominal quasi-static regime, potentially contributing to pressure fluctuations that are not directly associated with gas escape at the outlet. To be more specific, the bubble-velocity statistics show a clear porosity dependence. The cumulative distributions (CDF), shown in Figure 4 (b), indicate that lower porosity systems exhibit lower peak bubble velocities, consistent with stronger geometric confinement that limits rapid bubble mobilisation. The high porosity case exhibits the steepest fitted slope in Figure 4 (b), indicating that bubble velocity increases most strongly with bubble size. It also shows the clearest size-velocity correlation ($R_m = 0.69$). This relationship progressively weakens as porosity decreases, with $R_m$ declining to 0.48 for the medium porosity case and 0.28 for the low porosity case. In the low porosity packing, bubble velocity is therefore only weakly related to bubble size. The concurrent decreases in the fitted slope and correlation strength suggest that local pore-throat constraints increasingly govern bubble mobilisation, thereby reducing the influence of bubble size on translation velocity.

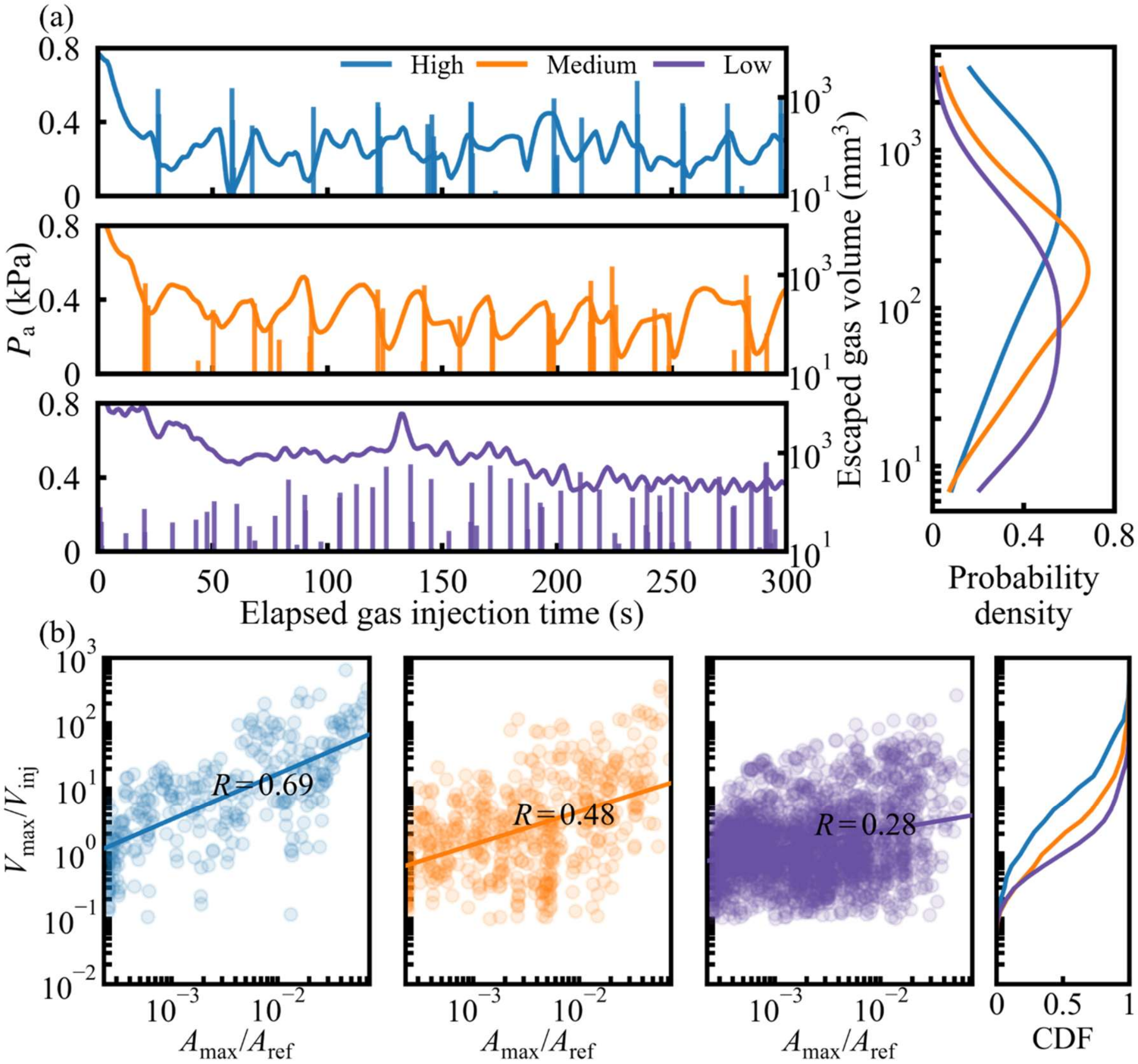

**Figure 4 Injection pressure evolution and gas escape dynamics.** (a) Apparent injection pressure with superimposed gas escape events during the first injection cycle (Air 1). The pressure is smoothed using a 5-s moving average, and escape magnitude is estimated from the observed escape area multiplied by the cell gap. The accompanying probability density distribution is evaluated using Gaussian kernel density estimation in log-transformed volume space; (b) Normalised maximum bubble velocity versus normalised bubble size.

## 3.2 Cyclic evolution of gas migration and pressure response

### *3.2.1 Overall trend of pressure evolution*

The temporal evolution of the injection pressure $P_{inj}^{*}$ over four successive gas injection cycles is shown in Figure 5. To emphasise cycle-to-cycle trends, raw signals were processed using a 60-second centred moving average. The low porosity case exhibits a clear reduction in mean injection pressure across cycles. This trend aligns with observations in reservoir-scale cyclic gas injection (Ma *et al*., 2016) and can be attributed to the evolution of flow connectivity (Yang *et al*., 2025). During initial invasion, the gas phase lacks established pathways, forcing the system to overcome high capillary entry pressures through intermittent breakthrough events. This phase is characterised by high-amplitude pressure oscillations. As the system undergoes repeated cycling, preferential flow channels progressively stabilize, allowing gas to bypass high-resistance zones. By the fourth cycle (15-20 min), the mean pressure converges toward a quasi-steady state, signalling that the establishment of continuous flow channels has effectively relieved local capillary constraints.

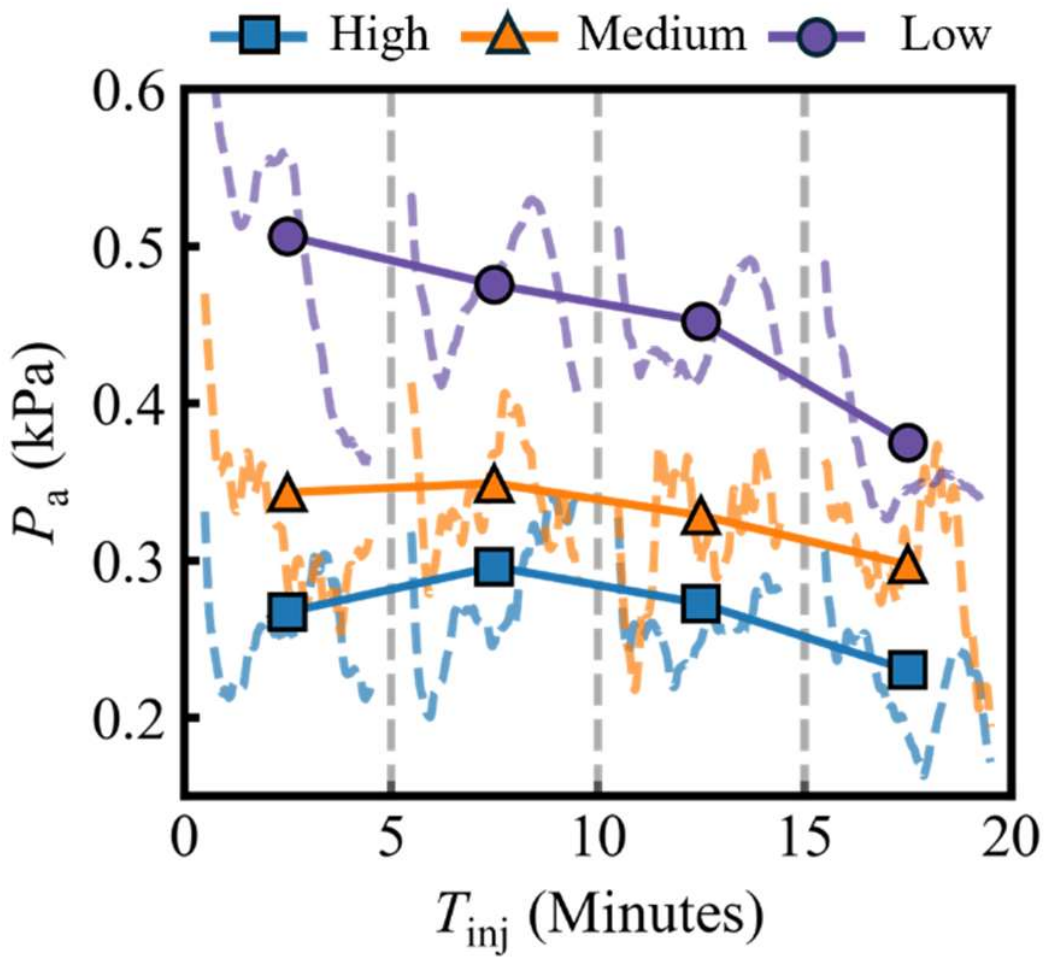

**Figure 5 Evolution of injection pressure during cyclic gas injection.** The dashed lines represent the smoothed values using a moving average with a window size of 60 seconds and the markers indicate the overall average apparent pressure calculated over consecutive 5-min intervals.

*3.2.2 Gas migration patterns and residual saturation*

In addition to the initial invasion behaviours, gas morphology continues to evolve throughout subsequent injection cycles. Figure 6 shows the gas distribution after alternative gas/water injection processes. Colours represent the size of gas clusters, which spans more than four orders of magnitude. The gas saturation $S_r$ is calculated as $S_r = \frac{\sum A_{G,i}}{L \cdot W \cdot \varphi}$, where $A_{G,i}$ is the projected gas area of the $i$-th gas cluster, $L$ and $W$ are the length and width of the Hele-Shaw cell domain, and $\varphi$ is the global porosity of porous media. Residual gas saturation was calculated as the average saturation during the final minute of each injection period. This window was chosen because both gas area and pressure were near quasi-stable over this interval, shorter windows were more sensitive to transient events, whereas longer windows could include early-cycle rearrangements. Figure 7 shows the ensemble mean of three replicates, with individual ones plotted in lighter colours. The trends are consistent across replicates, demonstrating good repeatability.

Relative to the first cycle, repeated drainage-imbibition increased residual trapped gas in all porosity conditions. The cycle-to-cycle increment decreased with cycle number, indicating an approach toward a steady trapping state. This general evolution and stabilisation of residual saturation over repeated injection cycles is consistent with prior studies (Herring *et al.*, 2016; Abdoulghafour *et al.*, 2020; Herring *et al.*, 2021; Ahn *et al.*, 2020; Wang *et al.*, 2023). Specifically, the high porosity system exhibits the most pronounced increase in residual gas trapping, reaching a terminal residual saturation more than 4 times that of its first cycle value. This increase coincided with a morphological transition from large connected cavities to more isolated blobs. Moreover, our results reveal the initial porosity also influences both the terminal residual saturation and the rate of convergence towards this steady state. The terminal residual gas saturation increases systematically with decreasing porosity, reaching 39%, 66%, and 80% for the high, medium, and low porosity systems, respectively. The rate of convergence to the terminal residual saturation, quantified by the ratio between its first-cycle and terminal residual saturation, also varies with porosity. This ratio is 23% for the high porosity case, 79% for the medium porosity case, and 83% for the low porosity case, indicating that denser packings reach their terminal state more quickly through cyclic injections.

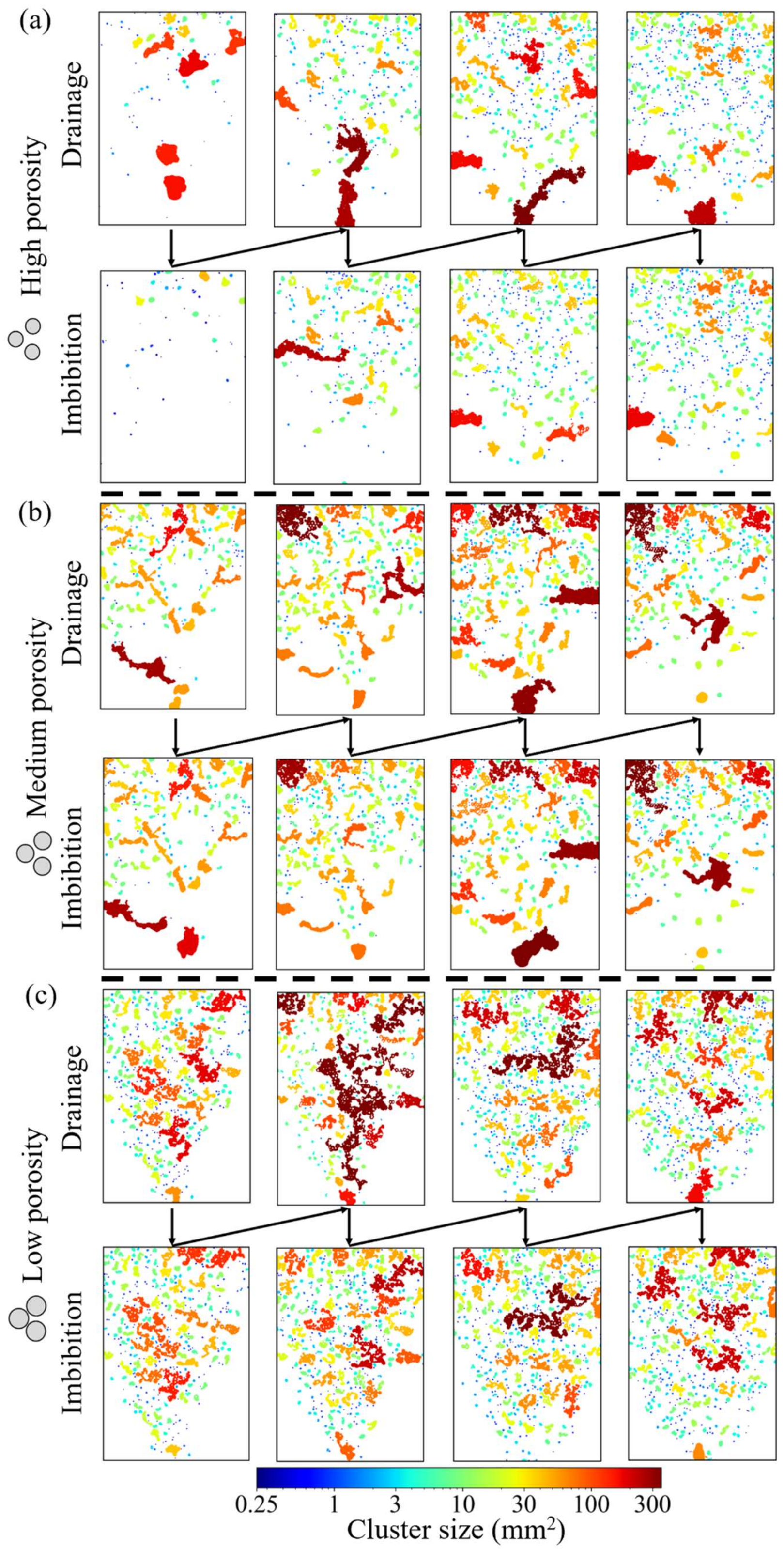

**Figure 6 Gas migration patterns during cyclic injections under different porosity conditions.** (a) High porosity; (b) Medium porosity; (c) Low porosity. The invading gas phase is visualised by cluster size, shown as a colour gradient from small clusters (0.25 mm², blue) to large clusters (350 mm², red).

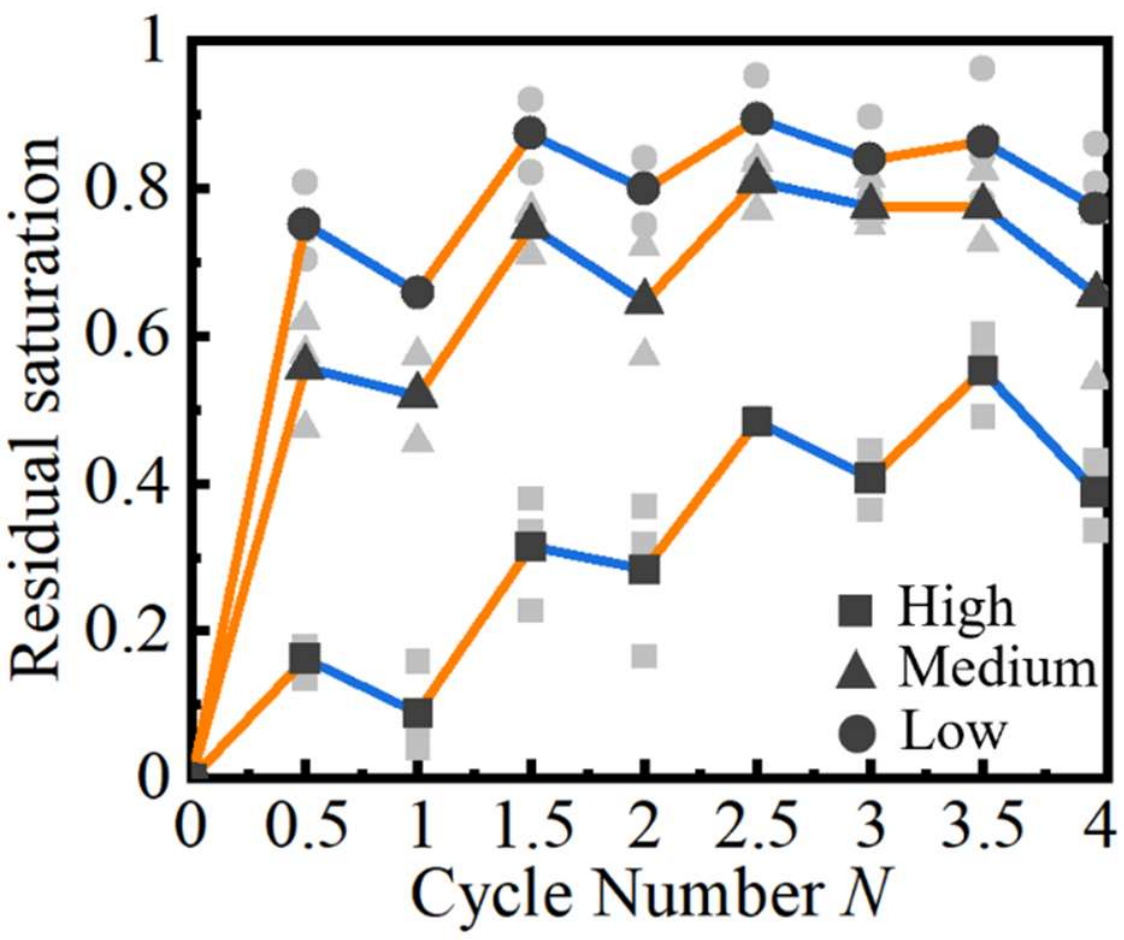

**Figure 7 Evolution of residual saturation**. Orange and blue lines represent drainage and imbibition processes, respectively. Each porosity condition includes three replicates, shown as shaded symbols.

### *3.2.3 Specific interfacial length*

To quantify the transition from cavity-like expansion to fingering-dominated migration, we characterised gas morphology using the specific interfacial length, defined as the ratio between the total fluid-fluid interfacial length $L_i$ and the projected gas-cluster area $A_{G,i}$ (Wang *et al.*, 2023). The analysis was performed over the final minute of each injection phase, with images sampled every 12s.

Figure 8 (a) plots the specific interfacial length against the residual gas saturation for systems with different initial porosities. A clear positive correlation is observed (with a coefficient of correlation, $R = 0.7$), indicating that higher residual saturation is generally associated with more interface-rich trapped gas structures. In addition, systems with lower initial porosity exhibit larger initial specific interfacial length. In the present system, effective porosity is influenced by two factors: the inherent porosity of initial packing and the progressive occupation of pore space by residual gas. Under this framework, lower effective porosity, either from lower initial porosity or increased gas trapping during cycling, tends to favour rougher water-air interfaces and more fragmented gas morphologies. These behaviours align with previous studies that with decreasing porosity, the invading pattern can shift from compact to fingering (Wang *et al.*, 2022).

To systematically quantify how specific interfacial length varies across porosities and evolves during repeated injection cycles, we applied kernel density estimation (KDE) contour fitting to characterise its statistical distribution as a function of individual cluster size. Figure 8 (b) shows an example of KDE contour for the high porosity case after gas injection in the first cycle. The KDE contours were plotted on a log-log scale and follow a power-law form $p(L/A_G) = c \cdot {A_G}^{-\alpha}$, where α is the geometric scaling exponent that quantifies the relation between the specific interface length and projected cluster size. The fitted curve exhibits excellent agreement with the data ($R^2 = 0.98$). Positive $\alpha$ values indicate that specific interfacial length decreases monotonically as cluster size increases, consistent with expected geometric trends. Notably, during the first gas injection, the high porosity case yields an exponent close to 0.5, matching the theoretical scaling expected for compact two-dimensional circular clusters ($L_i/A_{G,i} \sim {A_{G,i}}^{-1/2}$).

The evolution of the geometric scaling exponent $\alpha$ for systems with different initial porosity is plotted in Figure 8 (c), with the corresponding $R^2$ values indicated beside each data point. The black-solid line and black-dashed lines respectively represent the scaling exponent after the first (Air 1) and last (Air 4) cycle, respectively. All $R^2$ values exceed 0.9, justifying the power-law description across all experimental conditions. Overall, $\alpha$ decreases with initial porosity and increasing number of injection cycle, further deviating from the ideal scenario of circular clusters. These results highlight that interfacial morphology in deformable porous media not only depends on the initial structure but also evolves dynamically through coupled fluid-solid interactions during cyclic gas migration.

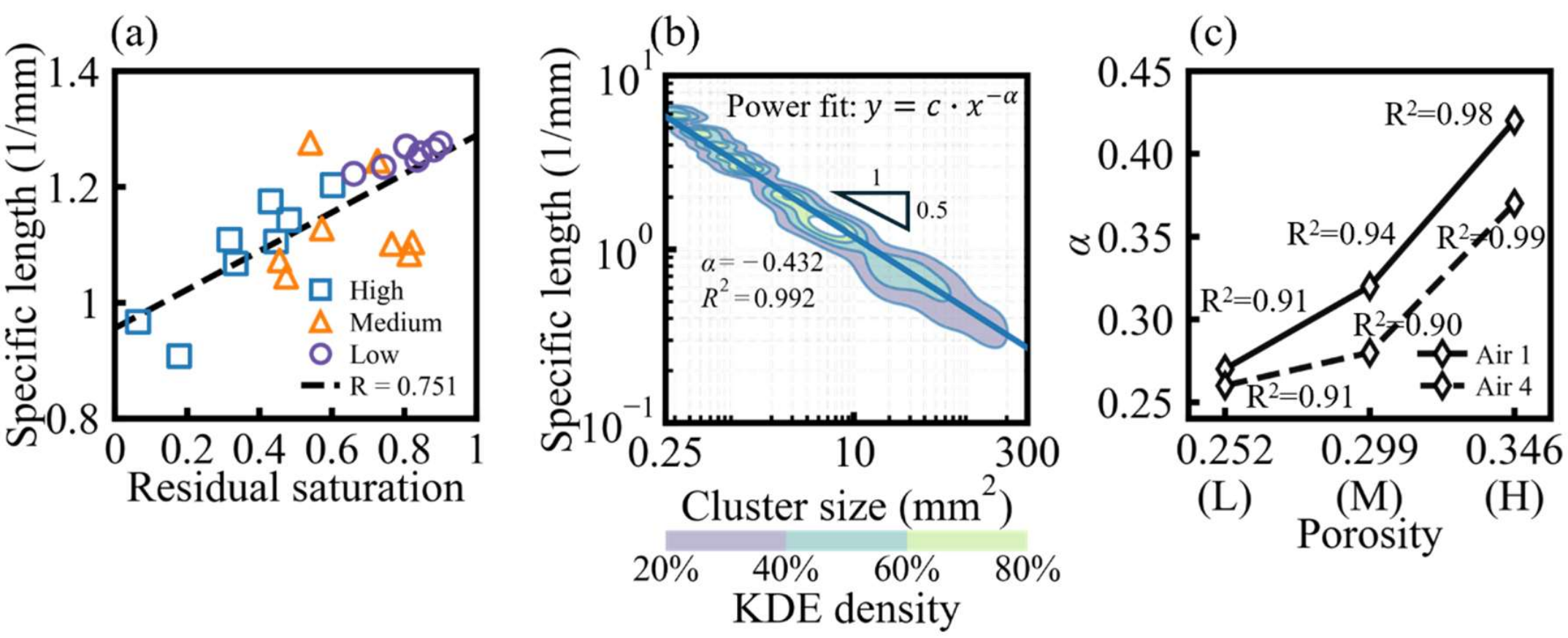

**Figure 8 Specific interfacial length characteristics**. (a) Overall specific interface length versus residual saturation. The black-dashed line is a linear fit with $R = 0.7$; (b) Statistical distribution of Specific interface length versus cluster size for cyclic injections for the 1st gas injection cycle for the high porosity test; (c) Geometric scaling exponent versus porosity after Air 1 and Air 4.

### 3.3 Discussion

This study employed a controlled laboratory environment in which air was injected into water-saturated deformable granular packings to understand the role of porosity in gas transport and trapping during cyclic injection. While these experiments provide valuable insights into gas migration dynamics, several aspects warrant further consideration to extend the applicability of the findings to practical engineering.

First, although the use of demineralised water and monodisperse soft particles provides a clean baseline for understanding fundamental mechanisms, real systems involve additional complexities, including salinity, wettability heterogeneity, and mineral dissolution-precipitation reactions (Al-Menhali *et al.*, 2015). Extending experiments to incorporate such factors would improve their practical relevance. In particular, the hydrogel particles are sensitive to temperature and salinity, suggesting that controlled variations in these conditions could be incorporated in future work and may offer valuable opportunities for studies with biological relevance (Mahon *et al.*, 2020).

Second, the present study investigates the effect of initial porosity under a fixed gravity-capillary condition (*i.e.*, identical Bond number). However, changing the Bond number can alter the relative importance of buoyancy and capillarity, thereby influencing upward gas migration, pathway destabilisation, local snap-off, and the redistribution or escape of disconnected gas clusters. Future work should systematically vary the Bond number, for example by changing cell inclination, density contrast, or interfacial tension, to determine how gravity-capillarity balance modifies gas invasion and trapping behaviours in deformable porous media.

# 4 Conclusions

This study investigated how initial packing porosity is associated with gas invasion, pressure response, and residual trapping in deformable porous media during cyclic injection. A vertical quasi-2D Hele-Shaw cell packed with soft hydrogel particles enabled synchronised pressure measurement and high-resolution visualisation of gas morphology over four drainage-imbibition cycles. The main findings are:

1. Initial gas entry into a water saturated hydrogel granular packing is primarily governed by capillary entry into the pore network. A Young-Laplace estimation based on an effective pore-throat width agrees reasonably with the measured entry pressures and captures the increase as porosity decreases. After entry, the invasion mode is dominated with porosity: decreasing porosity shifts the initial invasion mode from cavity dominated expansion toward localised pore invasion, with the medium porosity case showing a mixed mode characterised by cavity formation and front fingering.
2. Pressure evolution during cyclic injection reflects both pore-scale events and cycle-to-cycle gas distribution. Gas escape events trigger sharp pressure relaxations, with the fluctuation pattern depending on the frequency and volume of escaped gas. The low porosity case exhibits frequent small-amplitude pressure drops, whereas the high and medium porosity cases show more discrete pressure relaxations. Rapid internal gas redistribution may also contribute to the pressure response, as individual bubbles can translate at velocities more than two orders of magnitude higher than the nominal injection velocity, indicating localised departures from the nominally quasi-static regime. Across cycles, the overall decreasing mean pressure suggests preferential pathway reuse, which relieves local capillary constraints and lowers reinvasion pressure.
3. Residual gas saturation increases with cycle number and reaches higher terminal values as porosity decreases (*i.e.*, 39%, 66%, and 80% for the high, medium, and low porosity packings, respectively). The high porosity packing shows the largest relative cycle-to-cycle increase and the clearest morphological transition from connected cavities to isolated blobs, whereas the low porosity packing approaches its terminal saturation more rapidly.

4. Specific interfacial length reflects the available pore space, which is determined by the initial packing porosity and progressively reduced by residual gas occupation during cycling. Lower available pore space, either from lower initial porosity or greater residual gas trapping, favours rougher water-air interfaces and more fragmented gas morphologies. Kernel density estimation further shows that the cluster-scale specific interfacial length follows a power-law relationship with projected cluster area. Across all porosity levels, the geometric scaling exponent decreases with initial porosity and increasing injection cycles. These results demonstrate that interfacial morphology in deformable porous media depends not only on the initial packing structure but also on cyclically evolving gas trapping and fluid-solid interactions.

Overall, this study demonstrates that initial porosity is a key control on gas invasion, pressure response, and residual trapping in deformable porous media during cyclic injection. Lower porosity increases capillary constraints, promotes localised invasion and fragmented gas morphologies, and enhances residual retention, whereas higher porosity favours cavity expansion and larger release events. These findings suggest that porosity can regulate whether cyclic flow promotes fluid immobilisation or remobilisation in deformable pore networks, with implications for many engineering applications, such as underground gas storage and biological scaffolds. Future work should extend these observations to three-dimensional systems and systematically assess how physicochemical conditions, pore-scale heterogeneity, gravity-capillary effects, and coupled deformation modify gas trapping, redistribution, and release.

## Acknowledgements

The authors gratefully acknowledge Mr. Yue Zha from the School of Mechanical Engineering, University of Technology Sydney for providing valuable assistance during the testing phase. ZW acknowledges the financial supports from the Australian Research Council through project DE250100085.

## Availability Statement

The experimental images and image-processing code used in this study are openly available at https://github.com/HaiyiZ1/bubble_tracking. The repository includes the data and scripts required to reproduce the bubble segmentation, tracking, and post-processing analyses reported in this paper.

## Declaration of competing interest

The authors declare that they have no known competing financial interests or personal relationships that could have appeared to influence the work reported in this paper.

## Appendix A: Estimation of effective pore-throat width

The effective pore-throat width used in the Young-Laplace estimate was obtained by converting the measured particle volume and packing porosity into an equivalent projected spacing. For a representative particle volume $V_p$ and cell gap $b$, the in-plane cell area associated with one particle is:

$$A_{\text{cell}} = \frac{V_p}{b(1-\phi)}, \tag{A1}$$

where $\phi$ is the porosity, $V_p$ is the hydrogel volume, which can be calculated from the particle diameter $d_p$ as $V_p = \frac{\pi d_p^3}{6}$. Assuming a triangular arrangement of particle centres, the cell area is:

$$A_{\text{cell}} = \frac{\sqrt{3}}{2} s^2, \tag{A2}$$

where $s$ is the center-to-center spacing, giving

$$s = \left[\frac{V_p}{b(1-\phi)(\sqrt{3}/2)}\right]^{1/2}. \tag{A3}$$

For undeformed spherical particles, the in-plane throat width is $s - d_p$. However, this gives negative values for the present porosities, indicating that the soft hydrogel particles are deformed, flattened, or locally compressed. To obtain an effective hydraulic throat scale, the particle volume was converted to a volume-equivalent projected diameter, $d_{\text{eq}}$, by equating the spherical particle volume to the volume of an equivalent cylinder spanning the Hele-Shaw gap $V_p = \frac{\pi d_{\text{eq}}^2}{4} b$ (Jennings and Parslow 1988; Merkus, 2009), so that:

$$d_{\text{eq}} = \left(\frac{4V_p}{\pi b}\right)^{1/2}. \tag{A4}$$

The effective mean pore-throat width was then estimated as:

$$w_{\mathrm{eff}} = s - d_{\mathrm{eq}}. \tag{A5}$$